\documentclass[12pt]{article}
\usepackage{graphicx}
\usepackage{amsmath,latexsym,epsfig,epsf,rotate}

\begin{document}

\title{About Absoluteness of Data
on Elastic Electron Scattering with $^{12}$C Nucleus}
\author{ A.Yu. Buki\footnote{\normalfont
Corresponding author. E-mail address: abuki@ukr.net}, I.S.
Timchenko  \\
\emph{NSC ``Kharkov Institute of Physics and Technology'',} \\
\emph{Kharkov 61108, Ukraine } \\
}
\date{}
\maketitle

\begin{abstract}
The results obtained in Mainz in 1982 year were check up. The
analysis of data from this work was made at momentum transfer range
\mbox{$q = 0.25 \div 0.75\; fm^{-1}$} using the model independent
form factor (the expansion of form factor in a power series of
$q^2$) and the form factor corresponding to the distribution of
charge density in the shell model framework. We found a 3\%
systematical overestimation in Mainz data.
\end{abstract}

\vspace{0.2cm} PACS: 13.85.Dz

\vspace{0.2cm}

\section{Introduction}
\label{sec:introduction}

\hspace{0.5cm}
The results of electronuclear experiments are usually
reduced to absolute values by means of their normalization using
especially precise (master) data from elastic electron scattering.
These data are obtained from elastic electron scattering with
$^{12}$C or $^{1}$H nuclei, and, sometimes, with $^{4}$He nucleus.
During the experiment for the purpose of normalization in addition
to measurements with the nucleus under study we also measure elastic
electron scattering cross sections of one of the nuclei, for which
we possess reference data. The obtained cross sections are reduced
to the nucleus ground state form factor values $F_{el}(q_i)$. Using
the found values $F_{el}(q_i)$ we calculate the normalization factor
$$
K_i=\frac{F^2_{el,0}(q_i)}{F^2_{el}(q_i)},  \eqno(1)
$$
where $F_{el,0}(q_i)$ is the reference form factor; $q_i$ is the
momentum transferred to the nucleus.

The importance of the reference form factor precision in the
processing of experimental data was shown in work \cite {Fe}
performed in Darmstadt. Earlier the rms radii for $^{4}$He to
$^{209}$Bi nuclei (24 nuclei in all), normalized to measurements
with $^{12}$C nucleus from \cite {Be, End}, were obtained in this
laboratory. In view of uncertainties about the precision of data
from \cite {Be, End} new measurements of elastic electron scattering
cross section of $^{12}$C nucleus were carried out in Darmstadt.
Using this result the renormalization for all available data was
performed and the revised values of charge radii were obtained.

The latest and, obviously, the most precise work on elastic electron
scattering with $^{12}$C nucleus was carried out in Mainz lab \cite
{Re}. These data were used in the processing of our measurements
results. However, a question about the probability of a systematical
error in $^{12}$C nucleus data from ref.~\cite {Re} has arisen. The
present paper is dedicated to the study of the above mentioned
problem.

\section{Data analysis}
\label{sec:analysis}
\hspace{0.5cm}
 In Mainz lab the elastic electron scattering
cross section measurements of $^{12}$C nucleus were carried out at
\mbox{$q = 0.25 \div 2.75\; fm^{-1}$}. However, the measurements,
which the authors of this work consider absolute, were made at
\mbox{$q = 0.25 \div 0.75\; fm^{-1}$} (the rest of the measurement
results were relative and standardized to these data). Below we
shall only analyze data from the momentum transfer range \mbox{$q
\leq 0.75 fm^{-1}$}.

The data table of electron initial energies $E_0$, scattering angles
$ \theta$ and elastic electron scattering cross sections
$d\sigma/d\Omega $ measured on $^{12}$C nucleus can be found in
ref.~\cite{Re}. To use the results of this work for the normalization
procedure, it is necessary to find the squared form factor of
nucleus ground state $F^2_{el,0}(q_i)$ at different momenta transfer
$q_i$. For this purpose:
\begin{description}
\item[{\normalfont{(a)}}] let us transform $E_0$, $\theta$ and $d\sigma/d\Omega $
values to the corresponding values of  $F^2_{el,0}(q_i)$ and $q_i$;
\item[{\normalfont{(b)}}] let us select the analytical function  $F^2_{th}(q)$,
which will approximate the obtained $F^2_{el,0}(q_i)$ in the
momentum transfer range we are interested in. This is necessary to
avoid measuring the form factors $F_{el}$  at the same $q_i$ value
as reference form factors $F_{el,0}(q_i)$ during normalization using
eq.~(1).
\end{description}

Let us transform the $E_0$, $\theta$ and $d\sigma/d\Omega $ values
to the values of $F^2_{el,0}(q_i)$ and $q_i$, using well-known
formulas
$$
F^2_{el,0}=\frac{d\sigma/d\Omega}{\sigma_{Mott}}, \eqno(2)
$$
$$
q=\frac{2E_0}{\hbar c}\cdot
\frac{sin(\theta/2)}{\sqrt{\eta}}\cdot\xi, \eqno(3)
$$
where
$$
\sigma_{Mott}=\left(\frac{Ze^2}{2E_0}\right)^2\cdot\frac{cos^2(\theta/2)}{\eta\cdot
sin^4(\theta/2)}
$$
 is the scattering cross section on the nucleus with the charge
 number $Z$, $e$ is the electron charge;
$$
\eta=1+\frac{2E_0sin^2(\theta/2)}{M}
$$
is the kinematical correction, $M$ is the nucleus mass;
$$
\xi=1+\frac{3}{2}\cdot \frac{Ze^2}{\sqrt{5/3}\cdot <r^2>^{1/2} \cdot
E_0}
$$
is the correction, which takes into account the influence of the
nucleus Coulomb field on the incoming electron, $<r^2>$ is the
mean-square radius. Note that the formulas shown here are from
ref.~\cite{Re}.

For approximation of the obtained values $F^2_{el,0}(q_i)$  we use
simple presentations of nucleus ground state form factor
$F^2_{th}(q)$. As known, some of these presentations describe the
data at small momenta transfer \cite{Uberal} well enough and allow
to obtain the values of the rms radius with fairly good precision.
Such is the expansion of form factor in a power series of $q^2$,
which is
$$
F^2_{th}(q)=1-\frac{1}{3}\cdot a\cdot q^2 + \frac{1}{60}\cdot b\cdot
q^4 - \ldots\,, \eqno(4) 
$$
as well as the form factor of the nucleus ground state corresponding
to the distribution of charge density in the shell model framework.
For $^{12}$C nucleus this form factor can be expressed as follows
\cite{Gulkarov}
$$
F^2_{th}(q)=\left(1 - \frac{c^2\cdot q^2}{9}\right)^2 \cdot
exp\left(-\frac{d^2\cdot q^2}{2}\right).\eqno(5)
$$
Here $a$, $b$, $c$ and $d$ are parameters of fitting related to the
mean square radius: in the case of the form factor expansion in a
power series of $q^2$ (eq.~(4)) \mbox{$<r^2>=a$}, and for the form
factor with the distribution of charge density in the shell model
framework (eq.~(5)) \mbox{$<r^2>=\frac{2}{3}c^2+\frac{3}{2}d^2 $}.
Note that using of the eq.~(4) shows that the first three terms of
the series are enough
 for the approximation of the experimental form factors in studied range of
 q.

By definition \mbox{$\lim_{q\rightarrow 0} F^2_{el}(q)=1 .$} This
approach was used in some of the first $ee'$-scattering works and in
works with especially difficult conditions of measurements (for
instance, the measurements of electron scattering on $^3$H nuclei
implanted in titanium base \cite{Be_84}). Thus, a variable
multiplier $k$ was introduced in analytic presentation of form
factor which is fit to elastic electron scattering data. The $k$
value which was obtained as a result of the fitting is precisely the
normalization factor for absolutization of measured data. Using this
experience, we shall write the expression for the fitting function
as
$$
F^2(q)=k\cdot F^2_{th}(q).\eqno(6)
$$
If there is no systematic deviation in the data under study, it is
possible to assume the variable factor $k = 1.0$. Also, it is
possible to leave the $k$ factor as a variable parameter, however in
this case we have to obtain its value close to 1.0 within the limits
of the parameter errors.

The example of fitting eq.~(4) to Mainz data with and without
eq.~(6) is shown in fig.~1. The statistical precision of the data is
$0.45\%\div 0.49\%$ therefore the errors boundaries aren't visible
in the figure. The results of fitting the equations (4, 5, 6) to
these data are shown in table~\ref{tab:table}.

Since the value of the parameter $k$ appeared to be different from
1.0 approximately by 10 standard deviations, it is necessary to
check whether the obtained result is dependent on the analysis
conditions chosen. There are 16 experimental points in the examined
momentum transfer range, and among them there are two points for
each of \mbox{$q\,=\,0.25;\;0.35;\;0.45;\;0.55;\;0.74\;fm^{-1}$}.

\begin{figure}[th]
\label{fig1}
\begin{center}
\epsfig{file=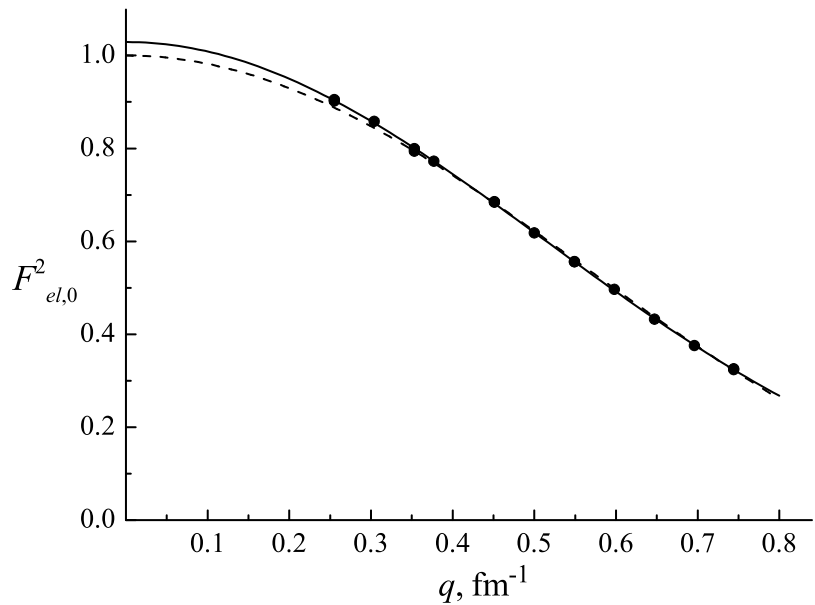,width=12cm} 
\end{center}
\caption{The squared form factor of $^{12}$C nucleus ground state.
The closed circles are the values obtained from the data of
ref.~\cite{Re}; solid line is the fitting of eq.~(4) with variable
parameter $k$ to these data; dashed line is the same fitting with
the fixed $k = 1.0$. }
\par
\end{figure}
\begin{figure}[th]
\label{fig2}
\begin{center}
\epsfig{file=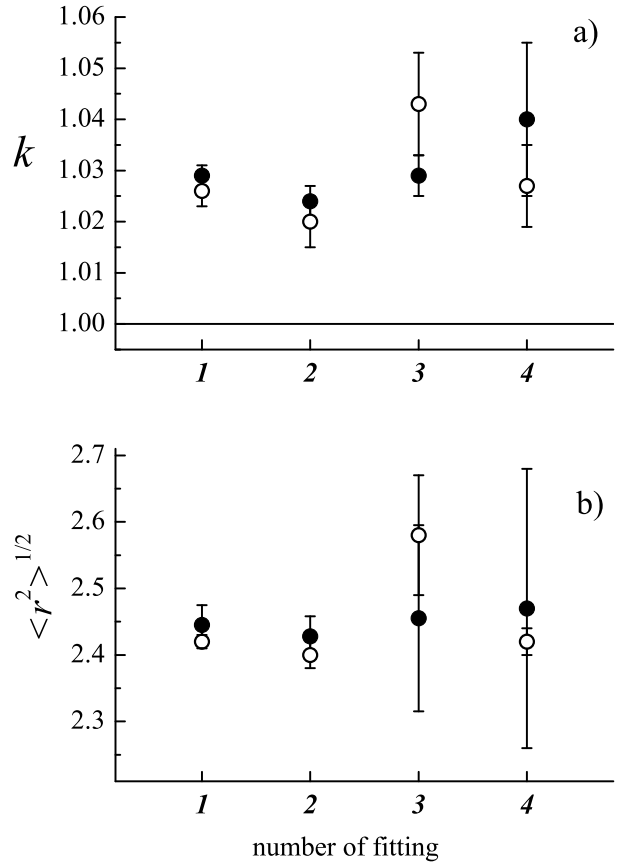,width=9cm} 
\end{center}
\caption{The results of the fittings of eq.~(6) with using eq.~(4)
(open circles)
 and with using eq.~(5) (close circles) to the different  ranges of data. The
horizontal scale the represents numbers of the fitting variants (see
text). a) $k$ is the normalization factor; b) $<r^2>^{1/2}$ is the
rms radius. }
\par
\end{figure}

\begin{table}[b]
\begin{center}
\caption{The result of fittings} \vspace{2mm}
\begin{tabular}{|c|c|c|c||c|c|c|}
\hline & $ k^{\ast)}$ & $<r^2>^{1/2}$&
$\chi^2_i$&$k$&$<r^2>^{1/2}$&$\chi^2_i$
\\\hline
power series in $q^2$&
1.0&$3.07\pm0.05$&962.0&$1.026\pm0.003$&$2.42\pm0.01$&0.71\\\hline
 shell model & 1.0 & $2.33\pm 0.03$& 5.0 & $1.029\pm 0.003$ &
$2.45\pm0.05$&0.75\\\hline
\end{tabular}
\label{tab:table}
\end{center}
\begin{flushright}
\vspace{-3.5mm} $\quad\quad\quad\quad$${\ast} ) $  The analysis with
fixed value $k = 1.0$ is shown in the left part of the table.
\end{flushright}
\end{table}

To verify whether the dependence of the obtained result on the
selection of fitting range is possible, we made a number of
fittings: \textbf{\textit{1}} -- all 16 points at \mbox{$q = 0.25
\div 0.75\;fm^{-1}$}; \textbf{\textit{2}} -- 13 points at \mbox{$q =
0.35 \div 0.75 \;fm^{-1}$}; \textbf{\textit{3}} -- 8 points at
\mbox{$q = 0.25 \div 0.45\; fm^{-1}$} and \textbf{\textit{4}} --  8
points at \mbox{$q = 0.50 \div 0.75 \;fm^{-1}$}. The results of this
analysis are shown in fig.~2.

\section{Discussion and Conclusion}
 \label{sec:DISCUSSION and CONCLUSIONS}
\hspace{0.5cm}
 First of all, it is necessary to note that in the
case of the fitting with the fixed value $k = 1.0$ we obtained the
improper $\chi^2_i$ ($ \chi^2$ per degree of freedom), while in the
case of the fitting with variable parameter $k$,  $\chi^2_i\approx
0.7$ (see table~\ref{tab:table}). As to the obtained values
$<r^2>^{1/2}$, within the limits of errors the identical values of
this magnitude were found   for two different presentations of form
factor (eq.~(4) and eq.~(5)) and variable $k$. The values
$<r^2>^{1/2}$ obtained in this case  are close to $2.456$ -- the
value of  the rms radius of $^{12}$C nucleus (this value is the
weighted mean of the results from a series of works \cite{Re,Ba}).
In case $k$ being fixed, there is considerable discrepancy in the
values of $<r^2>^{1/2}$.

Figure 2 shows that the values of variable multiplier $k$ and
$<r^2>^{1/2}$ which is obtained in this case within the limits of its
errors does not depend on the selection of the fitting range. Thus,
we can consider the existence of a systematical overestimation in
the data of ref.~\cite{Re} to be found. The  value of obtained
overestimation equals $2.6\% \div 2.9\%$, while the systematical
error declared in this work is $0.4\%$.

This conclusion should be taken into account using the data of work
\cite{Re} as master data.


\end{document}